\documentclass[final,3p,times,twocolumn]{elsarticle}

\usepackage[T1]{fontenc}

\usepackage{graphicx}

\usepackage{amsmath}

\usepackage{xspace}

\usepackage{amssymb}

\biboptions{comma,square}

\newcounter{bla}

\journal{Computer Physics Communications}

\newcommand{\Ane}{\ensuremath{\overline{\nu}_\mathsf{e}}}

\newcommand{\Ne}{\ensuremath{\nu_\mathsf{e}}}

\newcommand{\GeV}{\ensuremath{\,\mathrm{GeV}}}

\newcommand{\Dm}[1][42]{\ensuremath{\Delta m_{#1}^2}}

\newcommand{\ket}[1]{\ensuremath{\Bigl|{#1}\Bigr\rangle}}
\newcommand{\braket}[2]{\ensuremath{\bigl\langle{#1}\bigm|{#2}\bigr\rangle}}

\begin{document}

\begin{frontmatter}



\title{Calculation of oscillation probabilities of atmospheric neutrinos using nuCraft}


\author[a]{Marius Wallraff\corref{author}}
\author[a]{Christopher Wiebusch}

\cortext[author] {Corresponding author.\\\textit{E-mail address:} marius.wallraff@physik.rwth-aachen.de}
\address[a]{III. Physikalisches Institut, RWTH Aachen University, Germany}

\begin{abstract}

NuCraft (\texttt{nucraft.hepforge.org}) is an open-source Python project that calculates neutrino oscillation probabilities for neutrinos from cosmic-ray interactions in the atmosphere for their propagation through Earth. The solution is obtained by numerically solving the Schr\"odinger equation. The code supports arbitrary numbers of neutrino flavors including additional sterile neutrinos, CP violation, arbitrary mass hierarchies, matter effects with a configurable continuous Earth model, and takes into account the production height distribution of neutrinos in the Earth's atmosphere.

\end{abstract}


\end{frontmatter}



{\bf PROGRAM SUMMARY}

\begin{small}
\noindent
{\em Manuscript Title:} Calculation of oscillation probabilities of atmospheric neutrinos using nuCraft \\
{\em Authors:} Marius Wallraff, Christopher Wiebusch          \\
{\em Program Title:} nuCraft                                  \\
{\em Journal Reference:}                                      \\
{\em Catalogue identifier:}                                   \\
{\em Licensing provisions:} Revised BSD License               \\
{\em Programming language:} Python                            \\
{\em Computer:} IA32/x86-64 compatible                        \\
{\em Operating system:} all that are supported by SciPy, e.g., Linux, Windows, OS X \\
{\em RAM:} 134217728 bytes                                    
{\em Classification:} 1.1 Cosmic Rays, 11.1 General, High-Energy Physics and Computing,
	11.6 Phenomenological and Empirical Models and Theories \\
{\em External routines/libraries:} NumPy (1.5.1 or newer), SciPy (0.8.0 or newer)  \\
{\em Nature of problem:} Calculation of oscillation probabilities of neutrinos
	that originate in cosmic-ray interactions in the Earth's atmosphere and propagate
	through the Earth, for realistic Earth and atmospheric models and multiple flavors
	(optionally including sterile neutrinos and CP violation). \\
   \\
{\em Solution method:} Direct solution of the Schr\"odinger equation for $n$ flavors
	including matter effects, with sampling of the atmosphere.\\
   \\
{\em Restrictions:} Energy loss and absorption of neutrinos inside the Earth is
	not modeled; they have to be handled independently. \\
   \\
{\em Unusual features:} Completely configurable oscillation parameters (including
	optional sterile flavors), configurable and realistic Earth model including
	atmosphere. \\
   \\
{\em Running time:} Roughly 100 neutrinos per second and CPU core (depends on energy
	and oscillation parameters).\\
   \\

\end{small}

\section{Introduction}
\label{Intro}

Neutrino oscillations have been a major research topic for many particle and astroparticle physicists over the last decades \cite{Agashe:2014kda}. While neutrinos do not possess a mass in the minimum Standard Model of Particle Physics, many oscillation experiments have demonstrated that there are non-zero neutrino masses, and that their mass eigenstates differ from their flavor eigenstates. 
Despite the large progress that has been made in this field, there are still many open questions regarding neutrinos, including their absolute mass scale, their mass hierarchy, CP-violation, whether there are more than the three known flavors, and whether neutrinos are Majorana particles.
Additionally, some neutrino properties are not yet very well measured, and neutrino oscillations are a good phenomenon to improve our knowledge of those.

NuCraft is a Python project designed to compute oscillation probabilities of atmospheric neutrinos that originate from cosmic-ray interactions in the Earth's atmosphere. Many experiments are able to detect and measure atmospheric neutrinos. While several tools to calculate oscillation probabilities exist, e.g., {Prob3++} \cite{prob3,calland2014accelerated} or {GLoBES} \cite{Huber:2007ji}, nuCraft offers the advantage of a lightweight portable Python implementation and also improves the accuracy of calculations in some aspects. Beyond standard three-flavor oscillations, the code can handle an arbitary number of additional flavors, e.g., sterile neutrinos \cite{Abazajian:2012ys}.
A particular feature is the direct numerical solution of the Schr\"odinger equation with dynamically determined step sizes for the propagation through Earth.
Included is also the smearing of the propagation length caused by different production heights of neutrinos in the atmosphere. This improves the description of down-going and horizontally arriving neutrinos. Furthermore, the density profile of the Earth is not modeled by shells of constant matter density but is instead varied continuously.

\section{Theory}
\subsection{Neutrino oscillation in vacuum}

Neutrinos change their flavor during propagation in spacetime as a consequence of their weak-interaction flavor eigenstates $\nu_\alpha$ not being identical to their mass eigenstates $\nu_j$ \cite{kuo}; instead, they are a linear combination of each other, described by the unitary Pontecorvo-Maki-Nakagawa-Sakata (PMNS) matrix $U$:
\begin{align*}
	\nu_{\alpha} = \sum_{j} U_{\alpha j} \nu_{j}
\end{align*}

The PMNS matrix for $n \in \mathbb{N}$ neutrino flavors can be parameterized as a product of rotation matrices $R_{jk} \equiv R_{jk}(\theta_{jk}, \delta_{jk}) \in \mathbb{C}^{n \times n}$,
\begin{align*}
	U &= \prod_{j=1}^{n-1} \prod_{k=j+1}^{n} R_{jk}, \\
	{\left(R_{jk}\right)}_{mn} &= \hat{\delta}_{mn}\left(\left(\hat{\delta}_{jm}+\hat{\delta}_{km}\right) \cos(\theta_{jk}) + 1-\left(\hat{\delta}_{jm}+\hat{\delta}_{km}\right)\right) \\
	&\quad + \left(\hat{\delta}_{mj}\hat{\delta}_{nk}-\hat{\delta}_{mk}\hat{\delta}_{nj}\right)	\sin(\theta_{jk}) \exp \left((-1)^{\hat{\delta}_{nk}} \ i \ \delta_{jk}\right),
\end{align*}
with mixing angles $\theta_{jk} \in \mathbb{R}$, CP-violating Dirac phases $\delta_{jk} \in \left\{z \in \mathbb{C} \,\middle|\, |z|=1\right\}$, and Kronecker's delta $\hat{\delta}_{mn}$. This parametrization has to be given explicitly, because there is no clear canonical version for cases with $n > 3$, and rotation matrices do not commute in general. If neutrinos are Majorana particles, one also has to add CP-violating Majorana phases to the diagonal of the PMNS matrix, but those do not influence oscillation phenomena and are therefore omitted for this work. For antineutrinos, $U$ has to be replaced by its complex conjugate.

Neutrinos originate from weak interactions, so they are generated in definite flavor eigenstates. Their propagation through vacuum is described by the time-dependent $n$-dimensional Schr\"odinger equation ($c = \hbar = 1$)\footnote{Technically, as neutrinos are spin $1/2$-particles, one has to use a more general relativistic wave equation, but it can be shown that they all converge on what looks like a Schr\"odinger equation \cite{kuo}.}:
\begin{align*}
	i \frac{\mathrm{d}}{\mathrm{d}x}\ket{\nu_j} &= \widetilde{H}_0 \ket{\nu_j} \\
	\widetilde{H}_0 &= \operatorname{diag}\left(E_1, \ldots, E_n\right)
\end{align*}
In the ultra-relativistic limit, this is simplified using $E_j = \sqrt{p_j^2 + m_j^2} \approx E_\nu + \frac{m_j^2}{2 E_\nu}$, which implies $v = c$ and $p_j = p_1$ for all $j$. Both approximations are fulfilled well for atmospheric neutrinos, but they do not necessarily hold for neutrinos of cosmological origin, which are not the scope of this work. \\
Terms proportional to the identity matrix do not cause migration from one eigenstate into another and can be omitted; using the PMNS matrix to translate the Schr\"odinger equation into the flavor bases yields
\begin{align*}
	\frac{\mathrm{d}}{\mathrm{d}x}\ket{\nu_{\alpha}} &= \frac{-i}{2E_{\nu}} H_0 \ket{\nu_{\alpha}} \\
	H_0 &= U \operatorname{diag}\left(0,\Dm[21],\ldots,\Dm[n1]\right) U^\dagger.
\end{align*}
Transition probabilities can then be obtained by 
\begin{align}
	P_{\alpha\rightarrow\beta}(x) = \left| \braket{\nu_\beta(0)}{\nu_\alpha(x)} \right|^2. \label{eqProb}
\end{align}

\subsection{Matter effects} \label{ssecMatter}

When propagating through matter, neutrinos are subject to coherent forward scattering, which can strongly influence the oscillation behavior \cite{kuo}. The three known flavors of neutrinos can scatter on all particles via Neutral Current (NC) interactions, and $\Ne$ and $\Ane$ can additionally scatter via Charged Current (CC) interactions on electrons without being absorbed. In contrast, sterile neutrinos do not interact via weak interactions per definition. \\
These processes induce an effective squared mass:
\begin{align}
	\frac{\mathrm{d}}{\mathrm{d}x}\ket{\nu_{\alpha}} &= \frac{-i}{2E_{\nu}} \left(H_0+A\right) \ket{\nu_{\alpha}} \label{eqSchroed} \\
	A &= \operatorname{diag}\left( A_\mathrm{CC}+A_\mathrm{NC}, A_\mathrm{NC}, A_\mathrm{NC}, 0, \ldots \right) \nonumber \\
	A_\mathrm{CC} &= \pm 2 \sqrt{2}G_\mathrm{F} E_{\nu} Y_\mathrm{e} \rho / m_\mathrm{N}\nonumber \\
	A_\mathrm{NC} &= \mp \sqrt{2}G_\mathrm{F} E_{\nu} Y_\mathrm{n} \rho / m_\mathrm{N}, \nonumber
\end{align}
where $G_\mathrm{F}$ is Fermi's constant, $Y_\mathrm{e}$ is the electron fraction ($Y_\mathrm{e} = Y_\mathrm{p} = 1 - Y_\mathrm{n}$), $\rho$ is the mass density, and $m_\mathrm{N}$ is the mean of the proton mass and the neutron mass \cite{kuo}. The upper signs hold for particles, the lower for antiparticles. The reason for $A_\mathrm{NC}$ only to depend on the neutron density is that in electrically neutral and unpolarized matter, proton and electron potentials cancel out.

In some publications, matter effects are classified as either caused by the Mikheyev-Smirnov-Wolfenstein (MSW) effect or by parametric enhancement to gain phenomenological insights \cite{Parametric}. The work presented here correctly handles both, but the effects cannot be separated because they both originate from $A$ when numerically solving the Schr\"odinger equation.


\subsection{The interaction picture} \label{ssecInteractionPicture}

Numerical algorithms for solving ordinary differential equations generally work better the smoother the solutions are, such that the internal time steps can be chosen large without losing precision.
The solution for the Schr\"odinger equation (\ref{eqSchroed}) however is in most cases similar to the plane-wave solution for vacuum oscillations. To significantly reduce the number of time steps that the solvers need in these cases, the Schr\"odinger equation can be transformed into the interaction basis, in which the vacuum solution is a constant function \cite{InteractionPic}:
\begin{align}
	\frac{\mathrm{d}}{\mathrm{d}x}\ket{\nu_{\mathrm{inter}}} &= \frac{-i}{2E_{\nu}} \widetilde{A} \ket{\nu_{\mathrm{inter}}} \label{eqSchroedInt} \\
	\ket{\nu_{\mathrm{inter}}} &= \exp\left(-i H_0 x\right) \ket{\nu_{\alpha}}\nonumber \\
	\widetilde{A} &= \exp\left(-i H_0 x\right) \, A \, \exp\left(i H_0 x\right) \nonumber
\end{align}
The additional computations that are needed per time step are relatively expensive, but in most cases they are more than compensated for by the reduced number of steps required.

\section{Program}

NuCraft is fully written in Python and is compatible with both Python 2 (tested with 2.6 and above) and Python 3 (tested with 3.3). It relies on the libraries NumPy \cite{numpy} and SciPy \cite{scipy} and especially uses a SciPy wrapper around the ODE solver ZVODE \cite{VODE}. Using ZVODE, it directly solves the Schr\"odinger equation (\ref{eqSchroedInt}) in the interaction picture. The nuCraft source code is available at \cite{nuCraft} under the revised BSD license (3-clause version).

The project nuCraft consists of two classes, \texttt{NuCraft} and \texttt{EarthModel}. \texttt{NuCraft} is a class with three helper methods \texttt{ConstructMassMatrix}, \texttt{ConstructMixingMatrix} and \texttt{InteractionAlt} as well as the main methods \texttt{CalcWeights} and \texttt{CalcWeightsLegacy}, with the latter solving the Schr\"odinger equation in the original basis of equation (\ref{eqSchroed}). The legacy method does not perform the transformations described in subsection \ref{ssecInteractionPicture} and is therefore easier to read and faster in cases where vacuum oscillations are weak in comparison to matter effects (e.g., for sterile neutrinos at very high neutrino energies), but generally, it is substantially slower and does not offer all features, so its use is discouraged. \texttt{EarthModel} is an auxiliary class that allows for convenient and flexible specification of the parameters of the Earth that are relevant for oscillation effects of atmospheric neutrinos; see section \ref{sEarthModel}.

For information regarding the usage of the classes and methods, such as input parameters and output formats, please see docstrings and inline documentation.

\section{Earth Model \label{sEarthModel}}

\begin{figure}
	\includegraphics[width=\columnwidth]{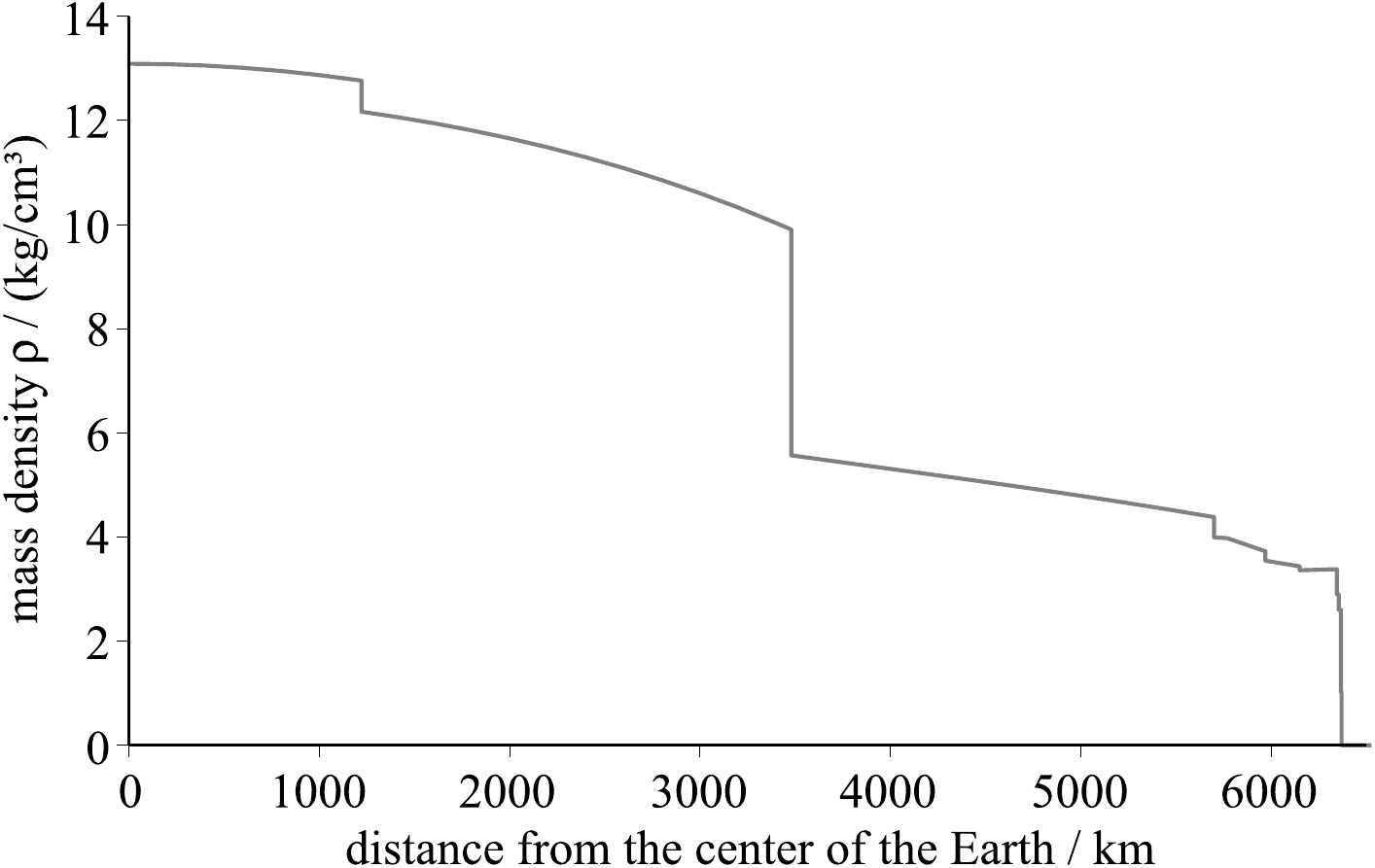}
	\caption{Mass density of the Earth as a function of the distance to Earth's center according to the Preliminary Reference Earth Model. The plot shows the values that are used in the code by default and without further simplifications.}
	\label{figPREM}
\end{figure}

As detailed in subsection \ref{ssecMatter}, the oscillation probabilities in matter depend on the parameters mass density $\rho$ and electron fraction $Y_\mathrm{e}$.
By default, nuCraft assumes the Earth to be spherical and uses the mass density values given by the Preliminary Reference Earth Model (PREM) \cite{prem}. The default electron fraction is $0.4957$ in the mantle (including the crust) and $0.4656$ in the inner and outer core.
The density profile is shown in figure \ref{figPREM}. It has been parameterized using 50 grid points that are interpolated by a linear spline.
During the solution of the Schr\"odinger equation, the minimizer step size is determined dynamically. Therefore, the density change between two steps is calculated quasi continuously.

The customization of the Earth model can be done by the aforementioned class \texttt{EarthModel}. 
Electron fractions in the three regions can be adjusted independently with a keyword argument, new density profiles can either be added to the dictionary \texttt{models} inside \texttt{EarthModel}, or can be loaded from a text file; an example file is provided with the code. Together with a trivial change in nuCraft's main class, this class can also be used to employ non-symmetrical Earth models, e.g., for use with reactor neutrino experiments; an explanation of the required changes can be found in the \texttt{README} file.

\section{Atmosphere}

Neutrinos are produced in the Earth's atmosphere at different heights.
For short neutrino path lengths and correspondingly shallow zenith angles, the variation of the production height becomes significant.
For atmospheric neutrinos in an experiment, the original production height is not known, so the oscillation path length will be smeared out by the distribution of production heights. As a result, the oscillation pattern becomes less pronounced.

\begin{figure}[t]
	\includegraphics[width=\columnwidth]{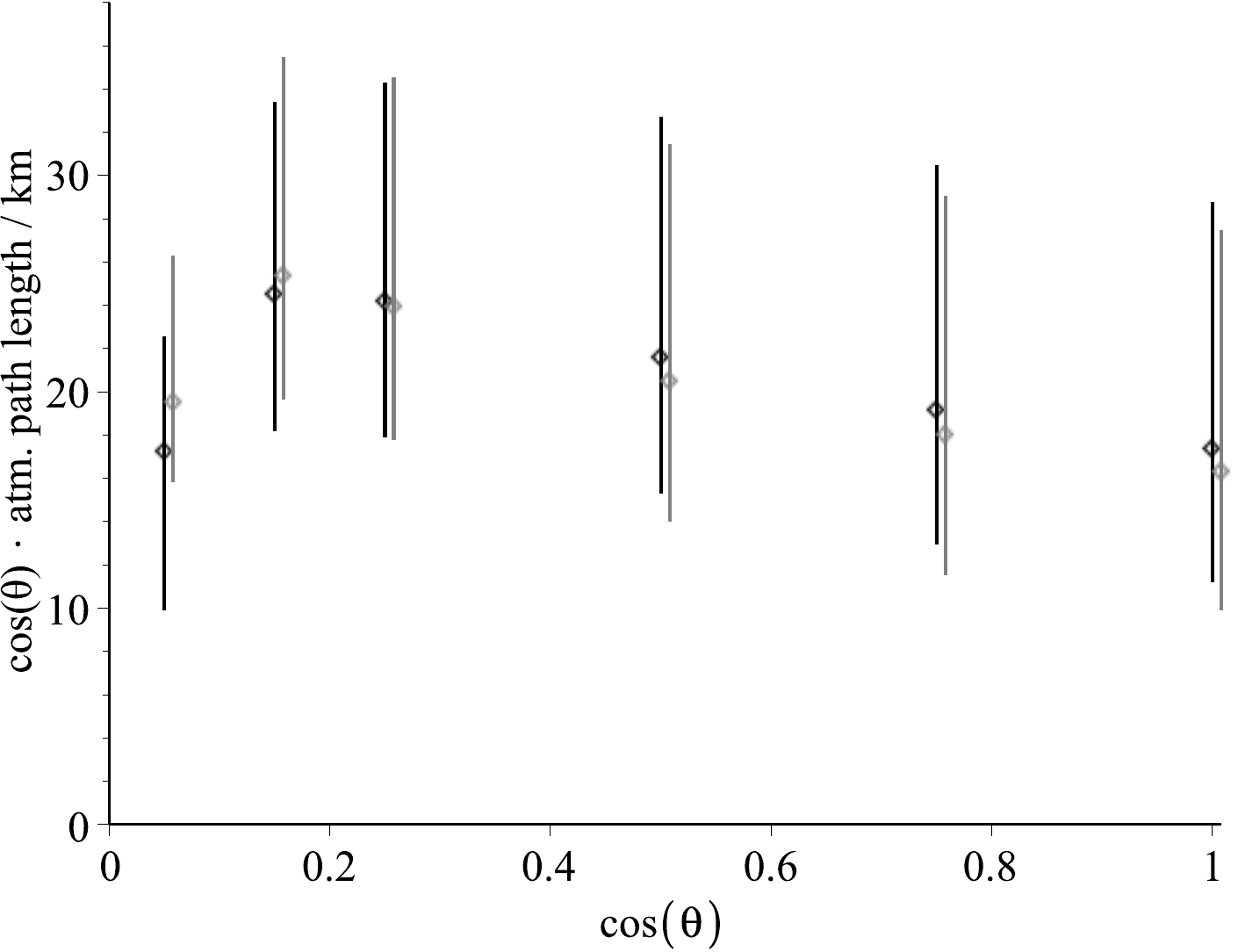}
	\caption{Expectation values and widths (defined as the $68\%$ probability range around the expectation value) of the neutrino path length in the atmosphere as function of the cosine of the zenith angle at two energies. Black corresponds to $200\GeV$, gray to $2\GeV$; gray has been shifted to the right by $0.008$.}
	\label{figAtmoWidths}
\end{figure}

\begin{figure}[thb]
	\includegraphics[width=\columnwidth]{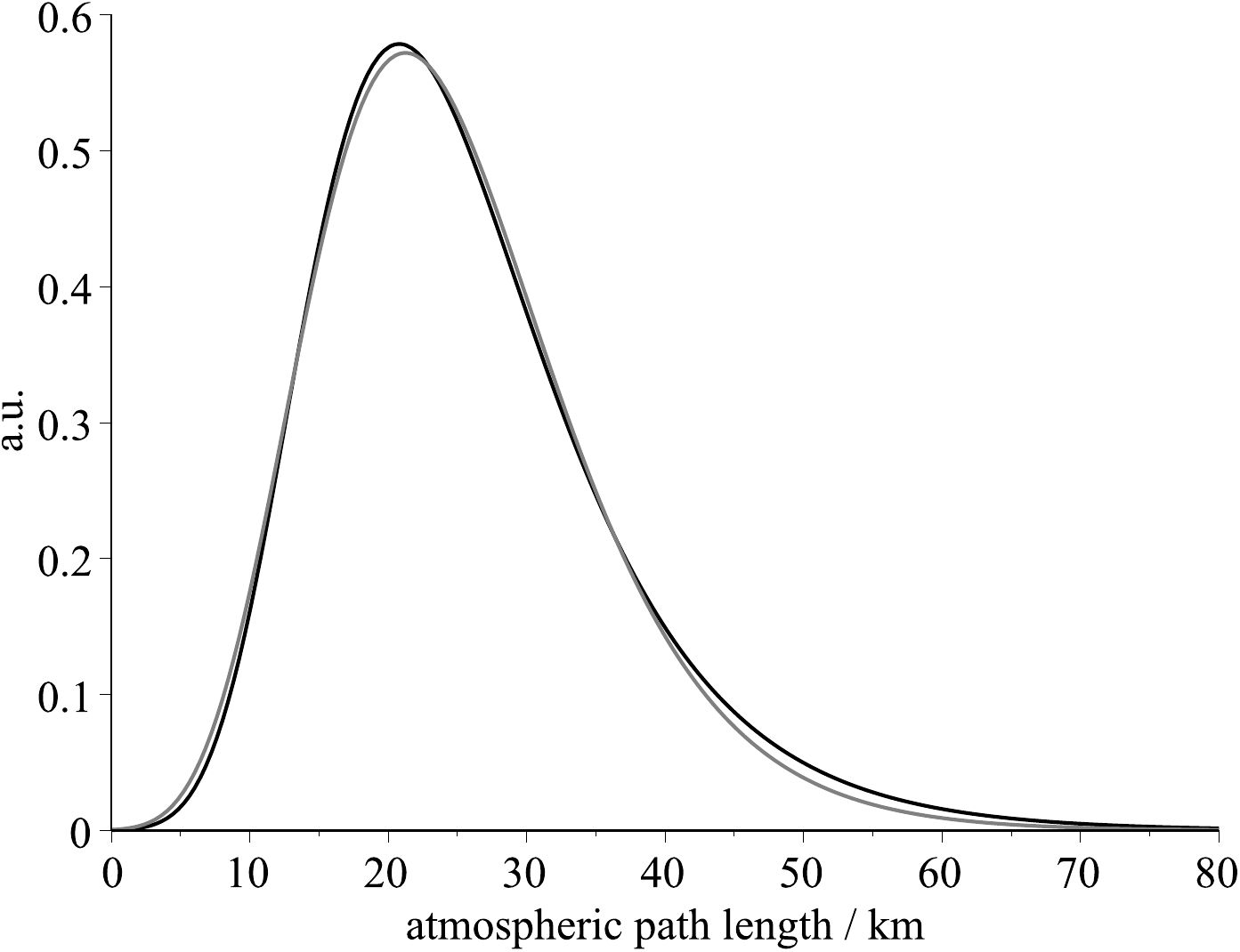}
	\caption{Exemplary unnormalized probability density function for the neutrino path length in the atmosphere at $2\GeV$ and $\cos(\theta) = 0.75$ (black; corresponding to a path length of $9556\,\mathrm{km}$ without atmosphere), and log-normal distribution fitted to it (gray).}
	\label{figAtmoLognorm}
\end{figure}

NuCraft uses the atmospheric model described in \cite{TomAltitude}. This model gives neutrino rates from meson and muon decays as functions of energy at six discrete zenith angle values, with no closed-form solution. 
The difference of the production heights between $2\GeV$ and $200\GeV$ is small compared to the width of their distributions (see figure \ref{figAtmoWidths}).
For the goal to model the dominant smearing effect, it was decided to fix the energy at $2\GeV$ for the calculation. The modified oscillation probability is not expected to depend strongly on the the specific shape of the distribution of production heights because the variance of heights over which the oscillation is averaged is large.

Based on the fit values in \cite{TomAltitude} the height distributions have been reproduced at each given zenith angle $\theta$. They are described reasonably well by log-normal distributions (figure \ref{figAtmoLognorm}) with two fitted parameters $\mu$ and $\sigma$.
To be able to interpolate to other zenith angle values, the fit results of these parameters were then parameterized as functions of the zenith angle, using a polynomial for $\mu$ and a power function plus a linear polynomial for $\sigma$.
Close to the horizon ($\left|\cos(\theta)\right| \lesssim 0.05$), \cite{TomAltitude} gives no reliable prediction. As the height distribution is up-down symmetric, the parameterization is smoothly interpolated  between up-going and down-going particles, using the value of $\left|\cos(\theta)\right|=0.05$ at the horizon.

By default, nuCraft computes eight equally likely production heights based on the quantile function of the log-normal parametrization.
Specifically, it uses the central values of the eight equally-sized subintervals of $\left[0,1\right]$.
The average oscillation probability for the eight heights is obtained efficiently: The oscillation probability is computed for the lowest height and then modified using analytically computed vacuum oscillation probabilities from the seven higher heights for the path length differences to the lowest height.
The chosen value of eight has been found as a good compromise between precision and speed and is sufficient for a reasonable smearing.
Alternative options are (1) production at a configurable fixed height, 
(2) a random height sampled from the continuous production height distribution,
or (3) to fully propagate eight neutrinos at eight equally likely production heights (see above) through Earth. The last option is slow and meant for debugging purposes only.

\section{Performance}

\begin{figure}[thbp]
	\includegraphics[width=\columnwidth]{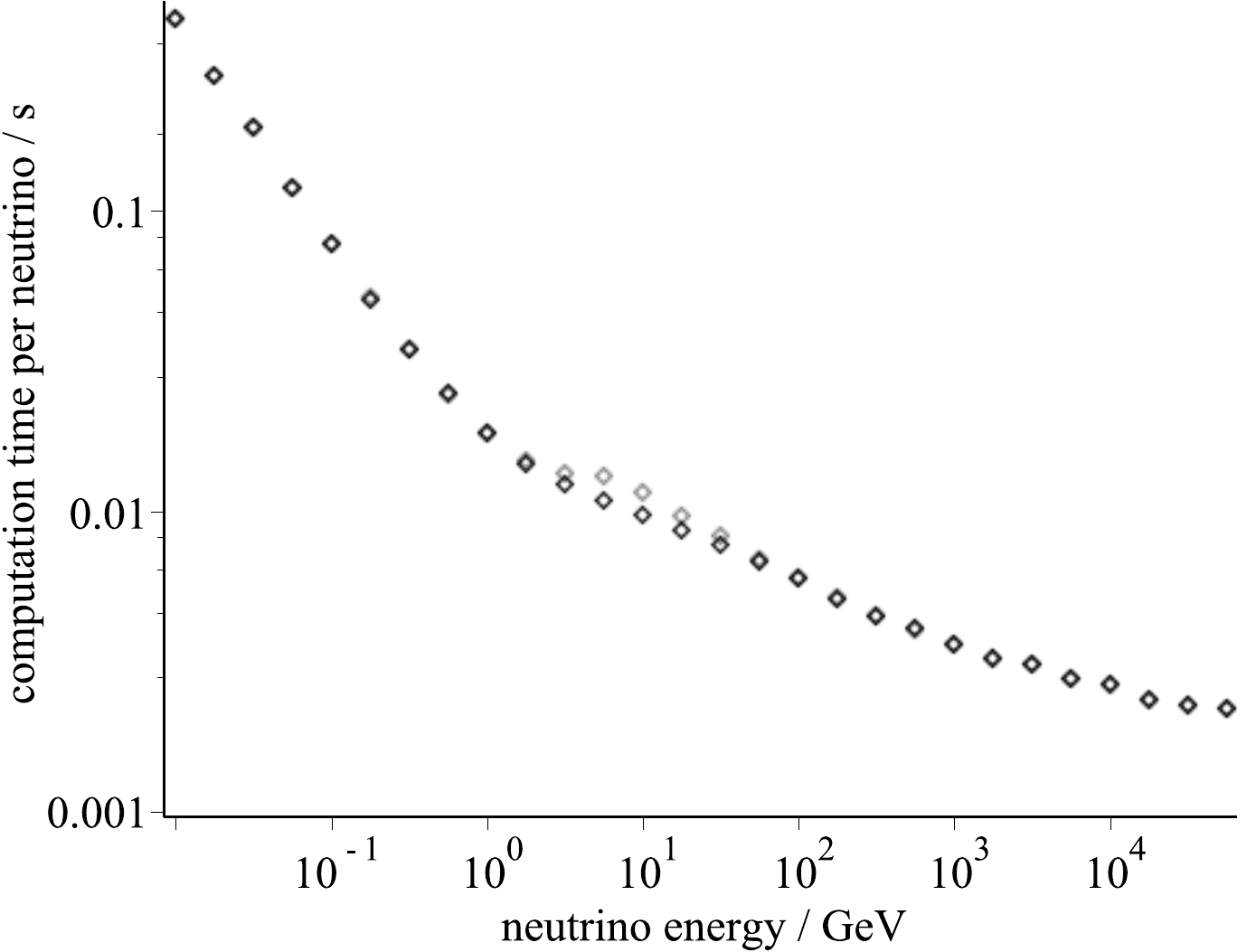}
	\caption{Time in seconds needed per oscillation probability computation in dependence of neutrino energy for the standard three-flavor model, for muon neutrinos (gray) and muon antineutrinos (black), distributed uniformly in $\cos(\theta)$. The feature at about $10\GeV$ is caused by matter effects. Times were measured using a single core of an AMD Phenom II X6 1055T CPU.}
	\label{figCPU}
\end{figure}

NuCraft prioritizes accuracy and flexibility over speed; moreover, Python is not the best choice for high-performance computing. Nonetheless, its speed can compete with similar tools written in C++ because of extensive use of highly optimized NumPy and SciPy functions. Figure \ref{figCPU} can be used to estimate the calculation speed. However, we note that the speed depends strongly on the neutrino energy, zenith angle, and the Earth model.

An intial comparison to Prob3++, written in C++, shows that nuCraft is about a factor 1000 slower. This large difference can be attributed to the Earth modeling because Prob3++ uses four layers of constant density, only. 
As the execution speed of Prob3++ scales linearly with the number of layers, an Earth modeling as accurate as nuCraft would lead to a substantialy reduced computing speed comparable to nuCraft.

\begin{figure}[thbp]
	\includegraphics[width=\columnwidth]{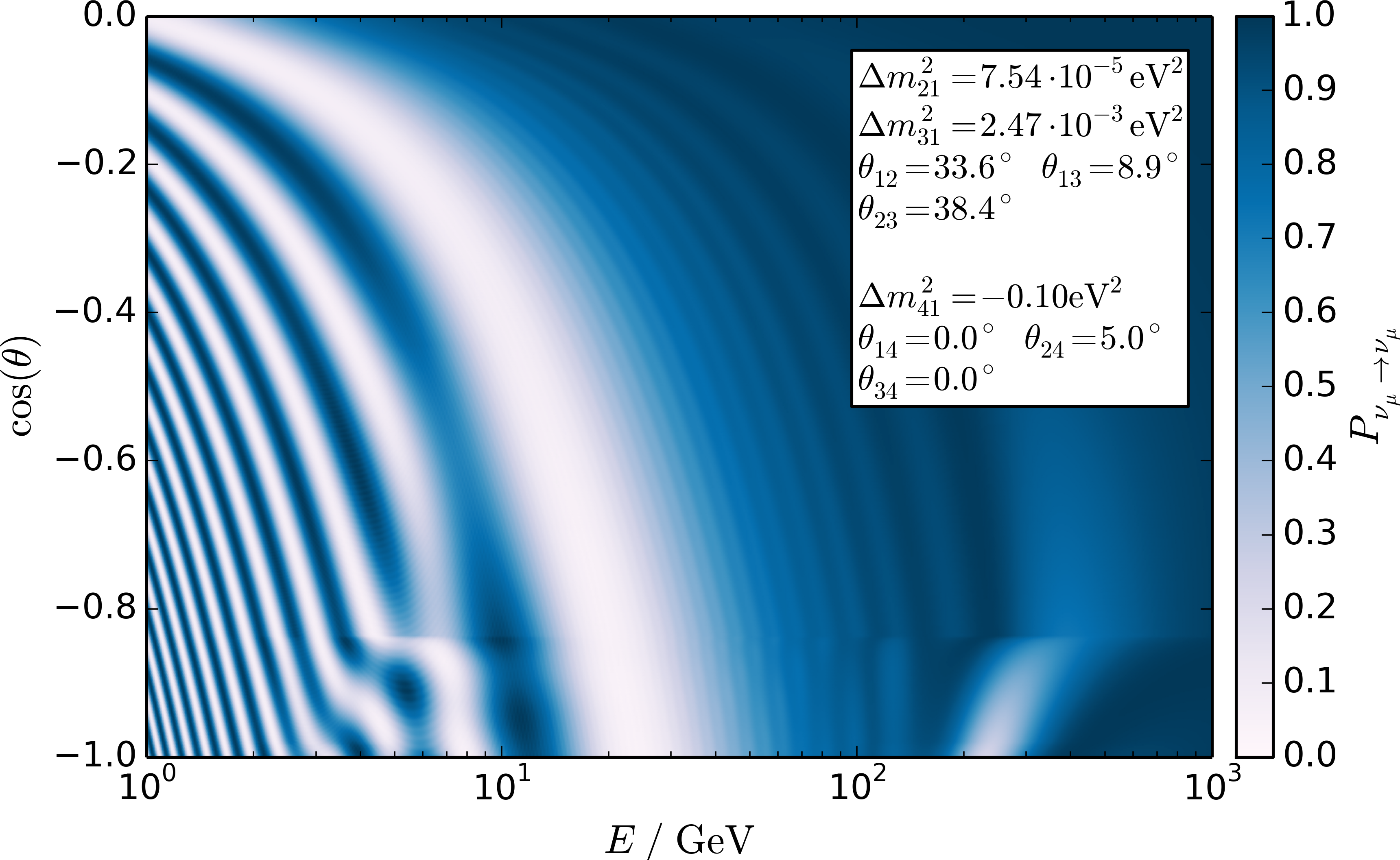}
	\caption{Example plot of muon neutrino disappearance probabilities computed with nuCraft in dependence of neutrino energy and zenith angle, including one sterile flavor. The hierarchy of the sterile neutrino was inverted ($\Dm<0$) to not have the matter resonances in the antiparticle channel. All relevant parameters are given in the plot.}
	\label{figOsci}
\end{figure}

Figure \ref{figOsci} shows an example oscillogram computed with nuCraft by calculating probabilities for one neutrino per grid point. In this example, one additional (sterile) neutrino flavor is added to the known three flavors.

\begin{figure}[thbb]
	\includegraphics[width=\columnwidth]{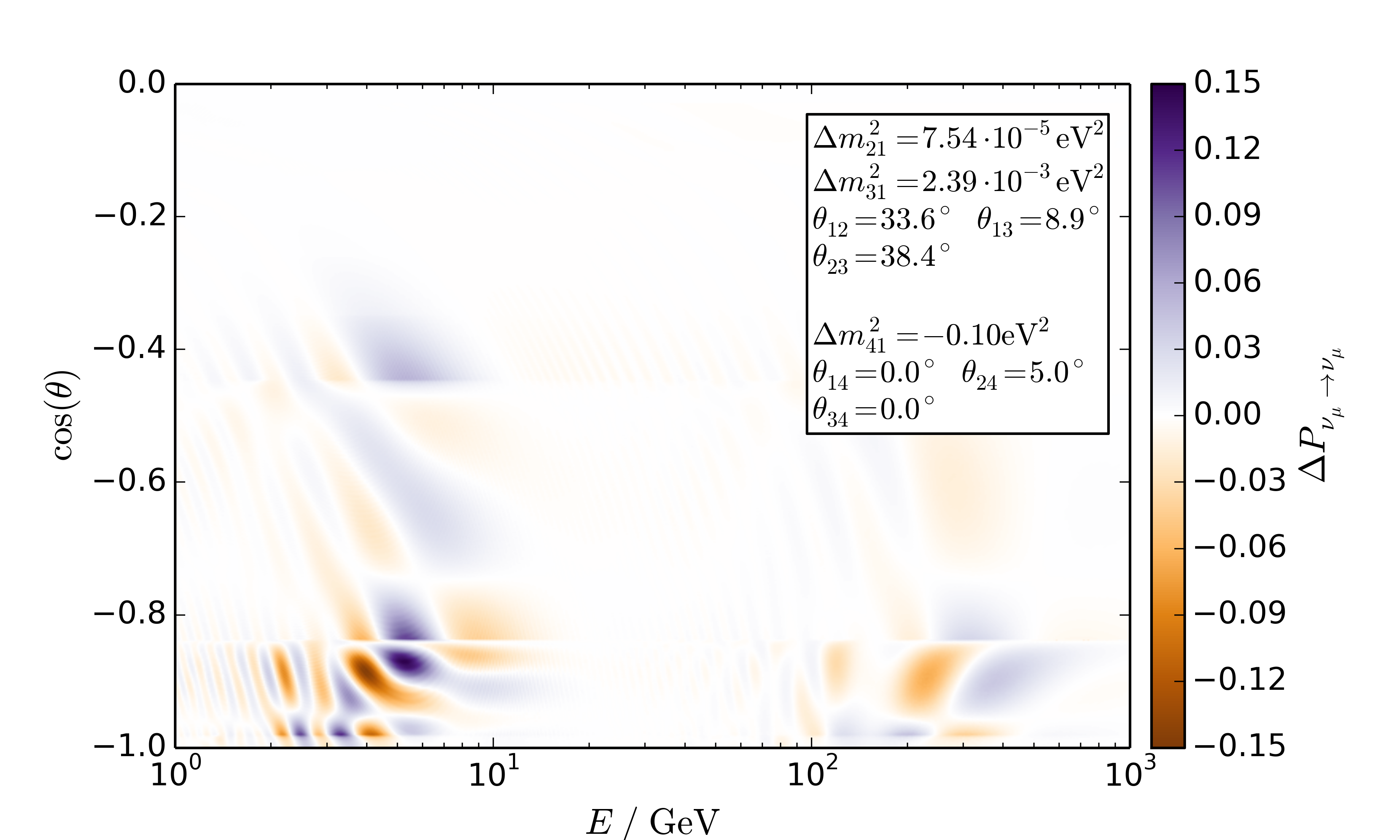}
	\caption{Difference of the oscillation probability for the default Earth model as implemented in nuCraft and an Earth model with 4 layers of constant density. All oscillation parameters are identical to fig.~\ref{figOsci}
 \label{figEarthImpact}}
\end{figure}

The effect of the Earth modeling is demonstrated in figure \ref{figEarthImpact}.
The more accurate description can lead to substantial differences up to 15\% in oscillation probabilities.

\begin{figure}[thbp]
	\includegraphics[width=\columnwidth]{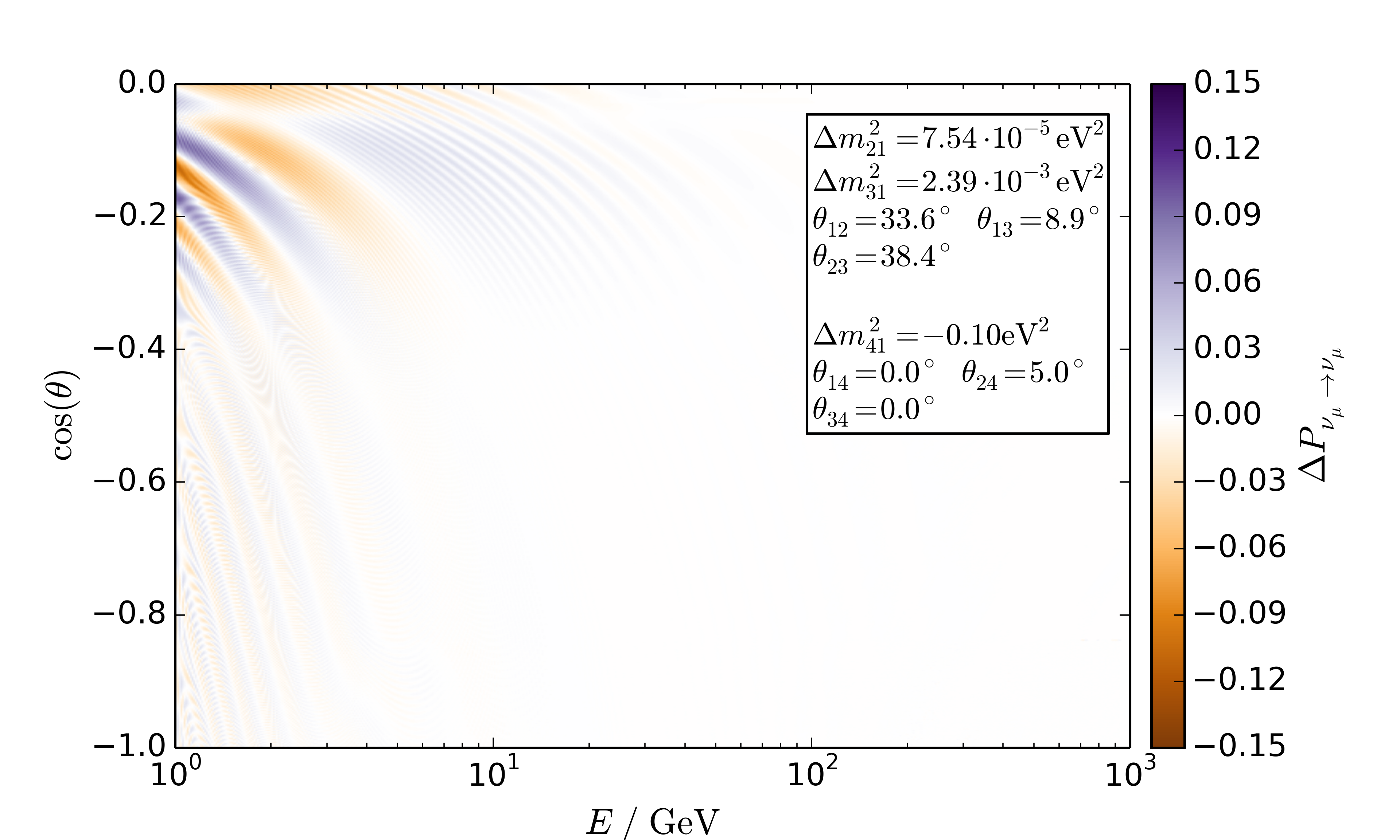}
  \caption{Difference of the oscillation probability for the modeling of the atmosphere as implemented in nuCraft to a fixed production height. All oscillation parameters are identical to fig.~\ref{figOsci} \label{figAtmImpact}}
\end{figure}

Figure \ref{figAtmImpact} shows the result of the smearing of the production height. Overall, the effect is small. However, close to the horizon, changes up to 10\% are visible.

The accuracy of the calculation is limited by the numerical solution of the Schr\"odinger equation. The accuracy is automatically estimated for each neutrino, using the sum of the survival and all oscillation probabilities to other flavors. This corresponds to a test of unitarity. For the default configuration the accuracy is better than $0.05$\%, which can be changed using the \texttt{numPrec} keyword argument.
Close to the horizon, further uncertainties arise from the modeling of the atmospheric production heights.

The neglection of the oblateness of the Earth only has small effects. The absolute error in length has its maximum at about $\cos(\theta) = -0.55$, but the more relevant relative error in the length barely exceeds $0.5\%$ at its peek at $\cos(\theta) = -0.1$ (using the WGS 84 model \cite{WGS84}). As oscillation probabilities scale with $\frac{L}{E}$, this error in length is substantially smaller than energy resolutions of typical experiments.

At very high energies, where neutrino interactions in the Earth become relevant for conventional neutrinos but not for sterile neutrinos, the propagation including sterile flavors can not be decoupled from oscillations.

\section*{Acknowledgments}

We would like to thank Tom Gaisser for his correspondence regarding the atmospheric model, Carlos Arg\"uelles for suggesting to use the interaction picture, the IceCube collaboration, and in particular Stefan Coenders, Denise Hellwig, Kai Krings, Martin Leuermann and Stefan Schoppmann, for critical feedback. This work was supported by the German Ministry of Education and Research (BMBF), the Deutsche Forschungsgemeinschaft (DFG), and the Helmholtz Alliance for Astroparticle Physics (HAP).





\bibliographystyle{elsarticle-num}
\bibliography{nuCraft-CPC}

\begin{thebibliography}{10}
\expandafter\ifx\csname url\endcsname\relax
  \def\url#1{\texttt{#1}}\fi
\expandafter\ifx\csname urlprefix\endcsname\relax\def\urlprefix{URL }\fi
\expandafter\ifx\csname href\endcsname\relax
  \def\href#1#2{#2} \def\path#1{#1}\fi

\bibitem{Agashe:2014kda}
K.~Olive, et~al.,
  \href{http://iopscience.iop.org/1674-1137/38/9/090001}{{Review of Particle
  Physics}}, Chin.Phys. C38 (2014) 090001.
\newblock \href {http://dx.doi.org/10.1088/1674-1137/38/9/090001}
  {\path{doi:10.1088/1674-1137/38/9/090001}}.
\newline\urlprefix\url{http://iopscience.iop.org/1674-1137/38/9/090001}

\bibitem{prob3}
R.~Wendell, \href{http://www.phy.duke.edu/\~raw22/public/Prob3++}{{Prob3++}
  software for computing three flavor neutrino oscillation probabilities}
  (2012--).
\newline\urlprefix\url{http://www.phy.duke.edu/\~raw22/public/Prob3++}

\bibitem{calland2014accelerated}
R.~Calland, A.~Kaboth, D.~Payne,
  \href{http://arxiv.org/abs/1311.7579}{Accelerated event-by-event neutrino
  oscillation reweighting with matter effects on a gpu}, Journal of
  Instrumentation 9~(04) (2014) P04016.
\newline\urlprefix\url{http://arxiv.org/abs/1311.7579}

\bibitem{Huber:2007ji}
P.~Huber, J.~Kopp, M.~Lindner, M.~Rolinec, W.~Winter,
  \href{http://arxiv.org/abs/hep-ph/0701187}{{New features in the simulation of
  neutrino oscillation experiments with GLoBES 3.0: General Long Baseline
  Experiment Simulator}}, Comput.Phys.Commun. 177 (2007) 432--438.
\newblock \href {http://dx.doi.org/10.1016/j.cpc.2007.05.004}
  {\path{doi:10.1016/j.cpc.2007.05.004}}.
\newline\urlprefix\url{http://arxiv.org/abs/hep-ph/0701187}

\bibitem{Abazajian:2012ys}
K.~Abazajian, M.~Acero, S.~Agarwalla, A.~Aguilar-Arevalo, C.~Albright, et~al.,
  \href{http://arxiv.org/abs/1204.5379}{{Light Sterile Neutrinos: A White
  Paper}}, {arXiv e-prints}.
\newline\urlprefix\url{http://arxiv.org/abs/1204.5379}

\bibitem{kuo}
T.~K. Kuo, J.~Pantaleone,
  \href{http://www.nikhef.nl/\textasciitilde{}h84/matterosc.pdf}{Neutrino
  oscillations in matter}, Rev. Mod. Phys. 61 (1989) 937--979.
\newblock \href {http://dx.doi.org/10.1103/RevModPhys.61.937}
  {\path{doi:10.1103/RevModPhys.61.937}}.
\newline\urlprefix\url{http://www.nikhef.nl/\textasciitilde{}h84/matterosc.pdf}

\bibitem{Parametric}
E.~K. Akhmedov, \href{http://arxiv.org/abs/hep-ph/9907435}{Parametric resonance
  in neutrino oscillations in matter}, Pramana 54 (2000) 47.
\newline\urlprefix\url{http://arxiv.org/abs/hep-ph/9907435}

\bibitem{InteractionPic}
C.~A. Arg{\"u}elles, J.~Kopp, \href{http://arxiv.org/abs/1202.3431}{{Sterile
  neutrinos and indirect dark matter searches in IceCube}}, JCAP 1207 (2012)
  016.
\newblock \href {http://dx.doi.org/10.1088/1475-7516/2012/07/016}
  {\path{doi:10.1088/1475-7516/2012/07/016}}.
\newline\urlprefix\url{http://arxiv.org/abs/1202.3431}

\bibitem{numpy}
D.~Ascher, P.~F. Dubois, K.~Hinsen, J.~Hugunin, T.~Oliphant, et~al.,
  \href{http://www.numpy.org/}{{NumPy}: Scientific computing with {Python}}
  (1995--).
\newline\urlprefix\url{http://www.numpy.org/}

\bibitem{scipy}
E.~Jones, T.~Oliphant, P.~Peterson, et~al.,
  \href{http://www.scipy.org/}{{SciPy}: Open source scientific tools for
  {Python}} (2001--).
\newline\urlprefix\url{http://www.scipy.org/}

\bibitem{VODE}
P.~N. Brown, G.~D. Byrne, A.~C. Hindmarsh,
  \href{http://dx.doi.org/10.1137/0910062}{Vode: a variable-coefficient ode
  solver}, SIAM J. Sci. Stat. Comput. 10~(5) (1989) 1038--1051.
\newblock \href {http://dx.doi.org/10.1137/0910062}
  {\path{doi:10.1137/0910062}}.
\newline\urlprefix\url{http://dx.doi.org/10.1137/0910062}

\bibitem{nuCraft}
M.~Wallraff, \href{http://nucraft.hepforge.org/}{{nuCraft} repository}
  (2013--).
\newline\urlprefix\url{http://nucraft.hepforge.org/}

\bibitem{prem}
A.~M. Dziewonski, D.~L. Anderson,
  \href{https://inspirehep.net/record/175521}{Preliminary reference earth
  model}, Physics of the Earth and Planetary Interiors 25~(4) (1981) 297 --
  356.
\newblock \href {http://dx.doi.org/10.1016/0031-9201(81)90046-7}
  {\path{doi:10.1016/0031-9201(81)90046-7}}.
\newline\urlprefix\url{https://inspirehep.net/record/175521}

\bibitem{TomAltitude}
T.~K. Gaisser, T.~Stanev, \href{http://arxiv.org/abs/astro-ph/9708146}{Path
  length distributions of atmospheric neutrinos}, Phys. Rev. D 57 (1998)
  1977--1982.
\newblock \href {http://dx.doi.org/10.1103/PhysRevD.57.1977}
  {\path{doi:10.1103/PhysRevD.57.1977}}.
\newline\urlprefix\url{http://arxiv.org/abs/astro-ph/9708146}

\bibitem{WGS84}
{National Imagery and Mapping Agency}, \href{http://earth-info.nga.mil/GandG/
  publications/tr8350.2/tr8350\_2.html}{{Department of Defense World Geodetic
  System 1984}}, Tech. Rep. TR 8350.2 Third Edition, Amendment 1, National
  Imagery and Mapping Agency (January 2000).
\newline\urlprefix\url{http://earth-info.nga.mil/GandG/
  publications/tr8350.2/tr8350\_2.html}

\end{thebibliography}







\end{document}